\documentclass{osa-article}
\journal{osac}
\articletype{Research Article}
\begin{document}

\title{Multislice forward modeling of Coherent Surface Scattering Imaging on surface and interfacial structures}

\author{Peco Myint\authormark{1,3}, Miaoqi Chu\authormark{1},Ashish Tripathi\authormark{1}, Michael J. Wojcik\authormark{1}, Jian Zhou\authormark{1,2}, Mathew J. Cherukara\authormark{1}, Suresh Narayanan\authormark{1}, Jin Wang\authormark{1}, and Zhang Jiang\authormark{1,4}}
\address{\authormark{1}X-ray Science Division, Advanced Photon Source, Argonne National Laboratory, Argonne, IL 60439, USA\\}
\address{\authormark{2}Center for Nanoscale Materials, Argonne National Laboratory, Argonne, IL 60439, USA\\
\authormark{3}pmyint@anl.gov\\
\authormark{4}zjiang@anl.gov\\
}

\begin{abstract}

To study nanostructures on substrates, surface-sensitive reflection-geometry scattering techniques such as grazing incident small angle x-ray scattering are commonly used to yield an averaged statistical structural information of the surface sample.  Grazing incidence geometry can probe the absolute three-dimensional structural morphology of the sample if a highly coherent beam is used. Coherent Surface Scattering Imaging (CSSI) is a powerful yet non-invasive technique similar to Coherent X-ray Diffractive Imaging (CDI) but performed at small angles and grazing-incidence reflection geometry. A challenge with CSSI is that conventional CDI reconstruction techniques cannot be directly applied to CSSI because the Fourier-transform-based forward models cannot reproduce the dynamical scattering phenomenon near the critical angle of total external reflection of the substrate-supported samples. To overcome this challenge, we have developed a multislice forward model which can successfully simulate the dynamical or multi-beam scattering generated from surface structures and the underlying substrate. The forward model is also demonstrated to be able to reconstruct an elongated 3D pattern from a single shot scattering image in the CSSI geometry through fast-performing CUDA-assisted PyTorch optimization with automatic differentiation. 
\end{abstract}

\section{Introduction}

With the advent of continually growing higher coherent flux form accelerator-based X-ray sources, it has become possible to resolve smaller spatial length scales at shorter experimental times, enabling new surface imaging techniques to emerge. Recent developments of state-of-the-art X-ray focusing techniques provides new insights into materials research \cite{sakdinawat2010nanoscale}. New techniques are being developed to take advantage of higher brilliance and coherence of X-rays to image or characterize surfaces at nano to sub-nano scales. For imaging small non-crystalline or heterogeneous structures on thick opaque substrates, coherent X-rays are essential to obtain surface scattering images that have rich spatial information — thus named Coherent Surface Scattering Imaging (CSSI) \cite{sun2012three}. Due to the nature of the detecting scattering, phase information is lost. Therefore, conventional coherent diffraction imaging methods resort to iterative forward and inverse Fourier transformations to reconstruct the real space information. However, these conventional phase retrival algorithms do not work straightforwardly for CSSI due to the dynamical scattering phenomenon of the grazing-incidence geometry. An appropriate model that can reproduce experimental scattering images in surface scattering geometry is therefore critical not only to explain experimental scattering patterns, but also to perform reconstructions. 

CSSI as a surface sensitive technique performed at the grazing-incidence geometry shares many characteristics as conventional grazing-incidence small-angle X-ray scattering (GISAXS) for probing nanostructures at surfaces or in thin films. The most significant phenomenon is the dynamical scattering effect which arises from the multiple scattering events of the photons due to the strong reflection from the substrate or film surface as well as the large X-ray illumination footprint. This phenomenon cannot be dealt with by single forward Fourier transform of the probed structures as adopted by the kinematic approximation for CDI analysis. A thorough understanding and the ability to reproduce the complex dynamical scattering patterns are necessary to extract accurate structural information. Distorted Wave Born Approximation (DWBA) is most often employed to deal with this situation\cite{Sinha1988, jiang2011waveguide}. In contrast to assuming a constant plane-wave like incident electric field in the kinematic approximation, the DWBA setting takes into account the strong substrate reflection (i.e. the major source of the distortion to the incident electric field) by means of a height-dependent electric field along the surface normal direction. Thus DWBA can only provide structural information statistically averaged in the sample surface plane rather than an absolute structure. In the presence of a heterogeneous structure of in-plane feature sizes comparable to the coherent length of the incident beam, the convenience of the in-plane statistical averaging in the DWBA's approximation vanishes. The accuracy of the DWBA deteriorates further if the electric field is largely distorted by the presence of heavy elements in the nanostructures. Hence arises the need to calculate the absolute location-dependent three-dimensional scattering field.

In this paper, we present an approach based on multislice to tackle the above challenges encountered in solving for nanostructures on surfaces and in thin films from coherent grazing-incidence scattering patterns. The multislice method naturally handles the three-dimensional potential field by means of iteratively producing the scattering from stacked two-dimensional slices and can accurately reproduce dynamical scattering effects. This is also done without assuming a close-form reciprocal-space shape factor as often adopted in the DWBA method. The original multislice method was originally proposed by Cowley and Moodie \cite{cowley1957scattering} to deal with the scattering of electrons by three-dimensional potential fields. It was later expanded by Goodman and Moodie \cite{goodman1974numerical} into numerical implementation on a computer. Multislice has long been used in acoustics and electron beam microscopy and has now been adapted to model the interaction of soft X-rays with an object of arbitrary shape and composition, given knowledge of its optical constants such as wavelength and beam profile. Multislice methods have been implemented in numerical calculations of optical interactions in electron \cite{smith1997realization,brown2020python} and X-ray \cite{ hare1994near, wang1998numerical, li2018multi} optics, but all of them were in transmission geometry. Recently, the multislice approach has be shown to be applicable to an even broader range of phenomena including X-ray reflectivity, where the substrate is part of the sample system. For example, Kenan Li et al. illustrated the total external reflection through multislicing \cite{li2017multislice}. Based on the aforementioned, our work sets to show that multislicing in grazing-incidence geometry can be used to study the 3D structure of substrate-supported surface patterns.

Our multislice formalism for reflection geometry requires that a three-dimensional object be created from which two-dimensional slices are made for wave to propagate through. Calculation of wave propagation through a single slice can be described by Equation \ref{equ:multislice}~\cite{paganin2006coherent}, where $k$, $n_\omega$, $\Delta$ and $\phi_\omega$ represent wavenumber of X-ray, refractive index, the Z (along $\hat{e}_Z$) thickness of a single slice, and the incoming complex wave probe.  By sequentially computing the exit wave of each slice using the exit waves of the previous slice, the final exit wave can be obtained. The final exit wave can then be Fourier transformed to obtain the far-field scattering pattern. This iterative procedure is illustrated in the Figure \ref{fig:schematic_geometry}a, where each slice is a projected two-dimensional object consisting of substrate, pattern, and air/vacuum integrated along $\hat{e}_Z$. 

Multislice method in transmission geometry has been implemented in high performing Graphic Processing Units (GPU) with CUDA for faster computational speed \cite{lobato2015multem}. The advantage of utilizing a GPU is that the forward calculations could be sped up to 60-100 times compared to only using a CPU. The prevalent usage of GPUs in machine learning has lead to the invention of the automatic differentiation tools such as PyTorch’s automatic differentiation engine\cite{ paszke2017automatic}, which is integrated into our multislice-assisted reconstructions, as discussed in Subsection \ref{subsec:CSSI-CDI structural refinement from experimental data: a 3D pattern created by Focused Ion Beam}. 3D samples with different materials (or refractive indexes) can be created in a GPU by voxelization of a pattern into 3D matrices with each element representing an integrated refractive index of materials which may be comprised of any desired materials. By containing multislice computations entirely within a GPU, one can manipulate 3D matrices such as rotations, scaling, and extending boundaries to emulate experimental conditions to perform realistic multislice forward calculations.

\begin{eqnarray}
 \phi_\omega(X,Y,Z=Z_0+\Delta)=&&\text{exp}(ik\Delta) \mathcal{F}^{-1} \Bigg\{ \text{exp}\Big(-i\Delta\frac{k_X^2+k_Y^2}{2k}\Big)\times \mathcal{F}\Bigg\{\nonumber\\&&\text{exp}\Big(\frac{k}{2i}  \int_{Z=Z_0}^{Z=Z_0+\Delta} [1-n_\omega^2(X,Y,Z)] \,dZ \Big)  \phi_\omega(X,Y,Z=Z_0)\Bigg\}\Bigg\} \nonumber\\
\label{equ:multislice}
\end{eqnarray}

Micro-scale 3D patterns used to obtain experimental and simulated CSSI images are listed in Table \ref{tab:samplelist}. These samples are smaller than the total illumination footprint of the incident probe. All of experimental data acquired from samples listed there are explained by the multislice forward model in Section \ref{sec:Fast Computing Multislice Simulations in GPU  and Experimental validation}. Reconstructions using the multislice model are demonstrated in Section \ref{sec:Reconstruction using the multislice forward model} in the form of Coherent Surface Scattering Imaging - Coherent Diffraction Imaging (CSSI-CDI). 

\begin{table}
\caption{\label{tab:samplelist}%
List of samples and specifications}
\centering
\begin{tabular}{cccc}
Sample Description&
\multicolumn{1}{c}{\textrm{Length ($\hat{z}$)}}&
\multicolumn{1}{c}{\textrm{width ($\hat{x}$)}}&
\multicolumn{1}{c}{\textrm{thickness ($\hat{y}$)}}\\
 & [$\mu$m] & [$\mu$m] & [nm] \vspace{1mm}\\
\hline 
\vspace{-3mm}
Elongated rod A ($\alpha_i$ = 0.5$^{\circ}$, $\psi$ = 0$^{\circ}$)& 70&4& 50 \\
\vspace{-3mm}
(Experiment and simulation)\\
Pattern material: Gold\\
\vspace{-3mm}
Elongated rod A ($\alpha_i$ = 0.5$^{\circ}$, $\psi$ = -0.5$^{\circ}$)& 70&4& 50 \\
\vspace{-3mm}
(Experiment and simulation)\\
Pattern material: Gold\\
\vspace{-3mm}
Buried 3D structure ($\alpha_i$ = 0.5$^{\circ}$, $\psi$ = 0$^{\circ}$) & 20 & 0.5 & 200 \\
\vspace{-3mm}
(Simulated reconstruction)\\
Pattern materials: Gold, Titanium, Silicon, Air \\
\vspace{-3mm}
FIB deposited 3D structure ($\alpha_i$ = 0.5$^{\circ}$, $\psi$ = 0$^{\circ}$) & 70 \& 70& 0.9 \& 0.3& 15 \& 10 \\
\vspace{-3mm}
Two layers on top of each other while centered\\
\vspace{-3mm}
(Experiment and simulation)\\
Pattern material: Platinum\\
\vspace{-3mm}
\end{tabular}
\end{table}

\section{Experimental Setups}

The experiments were conducted with a CSSI prototype at beamline 8-ID of the Advanced Photon Source, Argnonne National Laboratory, with 7.36 keV X-ray energy. The incident angles vary from around 0.3$^{\circ}$ to 1$^{\circ}$, where dynamical scattering is most pronounced at exit angles less than 0.6$^{\circ}$. It is important to also note that the critical angles of total external reflection of gold and silicon, two most common materials in our samples, determine the said dynamical scattering range. For experiments described here, the incident angle of 0.5$^{\circ}$ was chosen to observe pronounced dynamical scattering. The CSSI geometry enables X-ray penetration into our samples up to tens to hundreds of nanometers, yielding high sensitivity to depth and lateral spatial information. Our samples have either uniform or non-uniform depth profiles, through which it can be confirmed whether CSSI geometry experiments and multislice simulations have agreeing results. A coherent X-ray beam of $\sim$ 2$\mu$m $\times$ 2$\mu$m FWHM (measured from experiment, and can be approximated as a Gaussian profile) is used so the full sample resides in the illuminated area for CDI experiments, where our chosen CSSI geometry creates the X-ray footprints that sweep in the direction of the probe, ensuring the elongated samples are fully engulfed within the coherent X-ray beam. 

\begin{figure}[!ht]
\centering\includegraphics[width=13.3cm]{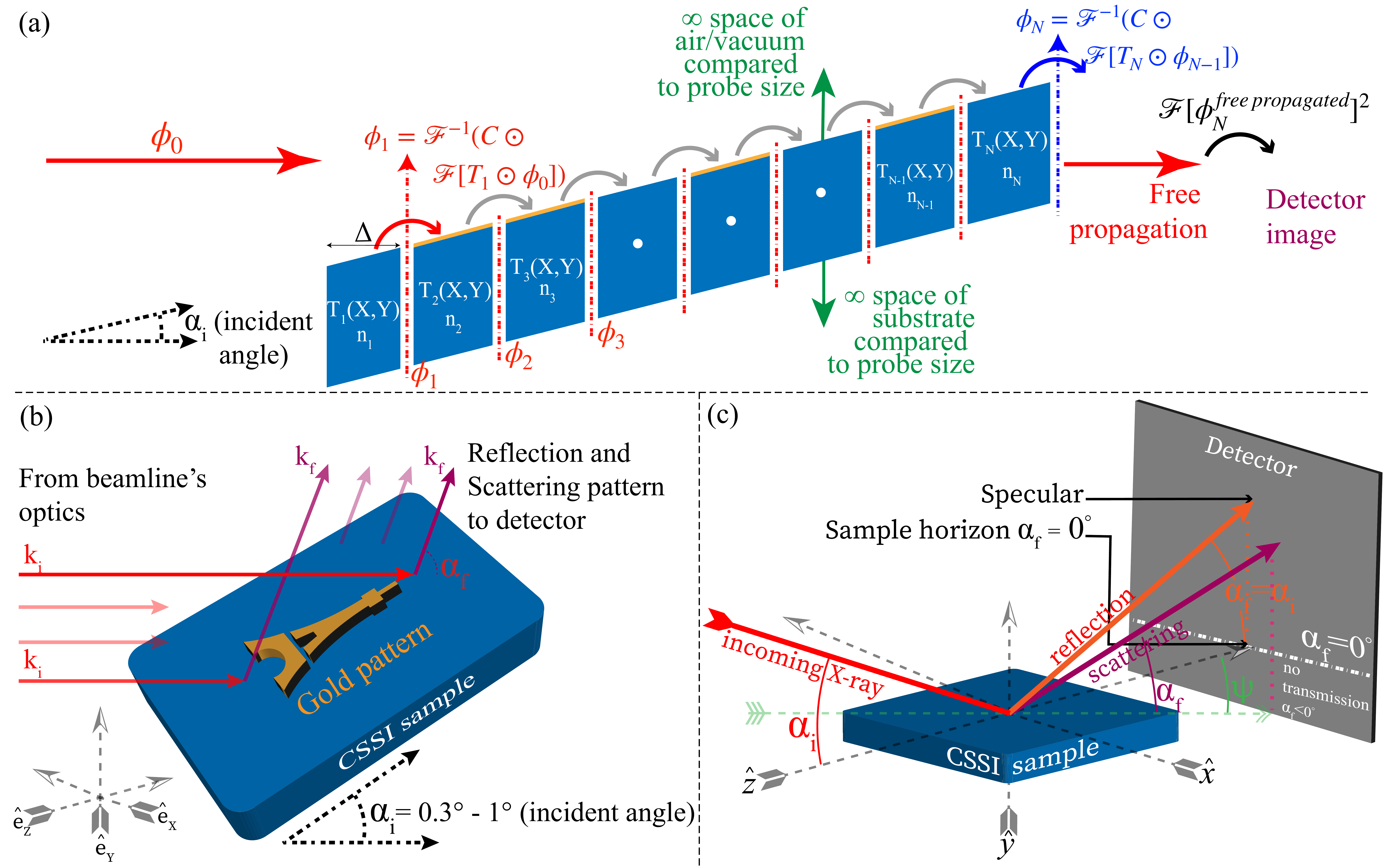}
\caption{Schematic diagrams of multislice and CSSI experimental setup: \textbf{(a)} Diagram explaining multislice which is also described by Equation \ref{equ:multislice}. $N$ is total number of $Z$ slices. $n$ is refractive index. $\Delta$ is a Z thickness for each slice. $T(X,Y)$ is a 2D transfer function in $\hat{e}_X$ and $\hat{e}_Y$ planes, integrated along $\hat{e}_Z$ for the thickness of $\Delta$, containing substrate, pattern, and air/vacuum. $C$ is a complex constant array containing information about the coordinates and X-ray energy. $\phi_0$ is an incoming/incident complex (with phase) X-ray probe along $\hat{e}_Z$, with wavenumber $k_i$. $\phi_1$, $\phi_2$, and $\phi_3$, $\phi_N$, are exit waves at slice 1, slice 2, slice 3, and slice N. The final detector image is the free propagated final exit wave $\phi_N^{\textrm{free propagated}}$ Fourier transformed and squared, thus the phase information is not present there as in experiments. \textbf{(b)} Experiment depicted in the lab frame of reference. \textbf{(c)} Experiment depicted in the sample frame of reference.}
\label{fig:schematic_geometry}
\end{figure}

The CSSI geometry is portrayed in two schematic diagrams in Figure \ref{fig:schematic_geometry}; (b) is the CSSI sample in the lab frame of reference (coordinates: $\hat{e}_X$, $\hat{e}_Y$, $\hat{e}_Z$) with an X-ray incident angle $\alpha_i$, whereas (c) is in the sample frame of reference (coordinates: $\hat{x}$, $\hat{y}$, $\hat{z}$) and $\alpha_{i}$, $\alpha_{f}$, and $\psi$ are X-ray incident angle, X-ray exit angle (in the $\hat{z}-\hat{y}$ plane), and X-ray exit angle (in $\hat{x}-\hat{y}$ plane). Equation \ref{equ:frameof_reference} summarizes how the index of refraction in the sample frame of reference is transformed into the lab frame of reference through yaw and pitch rotations, using the mentioned angles in rotation matrices described in Equation \ref{equ:rotation_matrix}.

\begin{eqnarray}
n_\omega(X,Y,Z) = R_{\textrm{pitch}}(\alpha_i )  R_{\textrm{yaw}}(\psi ) \; n_\omega(x,y,z)
\label{equ:frameof_reference}
\end{eqnarray}

\begin{eqnarray}
R_{\textrm{pitch}}(\alpha_i )={\begin{bmatrix}1&0&0\\0&\cos \alpha_i &-\sin \alpha_i \\[3pt]0&\sin \alpha_i &\cos \alpha_i \\[3pt]\end{bmatrix}},R_{\textrm{yaw}}(\psi )={\begin{bmatrix}\cos \psi &0&\sin \psi \\[3pt]0&1&0\\[3pt]-\sin \psi &0&\cos \psi \\\end{bmatrix}}
\label{equ:rotation_matrix}
\end{eqnarray}

The pattern with uniform height (the elongated rod A in Table \ref{tab:samplelist}) was made by e-beam lithography, whereas the non-uniform three-dimensional Platinum pattern (the FIB structure in Table \ref{tab:samplelist}) was made by Focused Ion Beam (FIB) deposition method. 

\section{Fast Computing Multislice Simulations in GPU and Experimental Validation}
\label{sec:Fast Computing Multislice Simulations in GPU  and Experimental validation}

The simulations are done in the lab frame of reference, meaning that the X-ray probe travels straight to meet an object that is tilted at a desired X-ray incident angle.  Simulations do not require high resolution to slice object along the direction of the probe because the small field of view (i.e. numerical aperture) along the $\hat{e}_Z$ direction in the grazing-incidence geometry. Higher resolution of the sample along the footprint direction can be achieved via in-plane tomography reconstruction which is not the focus of this paper and will be discussed elsewhere. The appropriate thickness $\Delta$ of an individual $Z$ slice (along $\hat{e}_Z$) and the sizes of the voxels in the lateral dimensions were inferred from the experimental numerical apertures, which correspond to the maximum achieved exit wave vector $k_{f}$ given by the flux in the experiment. The X-ray footprint determines how many total number of slices we need for the entire simulation.  Naturally, finer slice thickness only leads to longer computational times without necessarily yielding finer 3D electric field needed for the final scattering pattern; keeping 200 nm thickness per a $Z$ slice is observed to be a good choice for speed and accuracy for our working X-ray incident angles; in other words, selecting smaller $\Delta \;<$ 200 nm does not yield a significantly different scattering image, but only prolongs the simulation time. The lateral dimensions (the field view plane perpendicular to the direction of probe) of a voxel define the finest angular resolution the simulation can provide. To reproduce the same resolution and wavenumber range achieved in experiments, a minimum of 5 nm lengthscale or 0.2 nm$^{-1}$ wavenumber per voxel is required meaning that a significant 3D matrix size of refractive indexes is necessary to include pattern, substrate, air, and enough wave propagation distance. This is a memory-taxing and time-consuming computation and manipulation of the enormous 3D matrix can be quickly performed through a high-performing GPU code, if its memory can hold the matrix. 

CuPy\cite{cupy_learningsys2017} and PyTorch\cite{paszke2019pytorch} are used as fast-performing GPU computation tools to do object rotations and matrix multiplications required to do reflection geometry multislice wave propagation. For mere comparison between experimental data and forward calculation simulations of large samples discussed in this section, CuPy code was written in a memory efficient way to accomplish the task. For 3D object reconstruction purposes in Section \ref{sec:Reconstruction using the multislice forward model}, every calculation step has to be tracked for automatic differentiation, which is required to obtain gradients with respect to parameters, and PyTorch is therefore preferred albeit more memory taxing. For both CuPy and PyTorch GPU codes, from object creation of any desired 3D pattern of 70 $\mu$m (length, $\hat{z}$) $\times$ 4 $\mu$m (width, $\hat{x}$) $\times$ 50 nm (depth, $\hat{y}$) dimensions, along with substrate and air, to computing the final scattering intensity pattern, it takes approximately a second (on HPE NVIDIA A100 Graphic Card with 80 GB memory), enabling fast multiple iterations with different object guesses: a process intrinsic in reconstruction algorithms as discussed in Section \ref{sec:Reconstruction using the multislice forward model}. 

Although multislice is computationally more expensive than DWBA, it has advantages in that it enables finer three dimensional control of sample's compositions and yields accurate final scattering patterns. To test whether the multislice forward model can reproduce experimental data, the rod A pattern in Table \ref{tab:samplelist}, oriented in two positions, is used. The angles $\alpha$ and $\psi$ are calculated for simulated scattering pattern and are compared with the experimental data angle to angle.

\begin{figure}[!ht]
\centering\includegraphics[width=13.2cm]{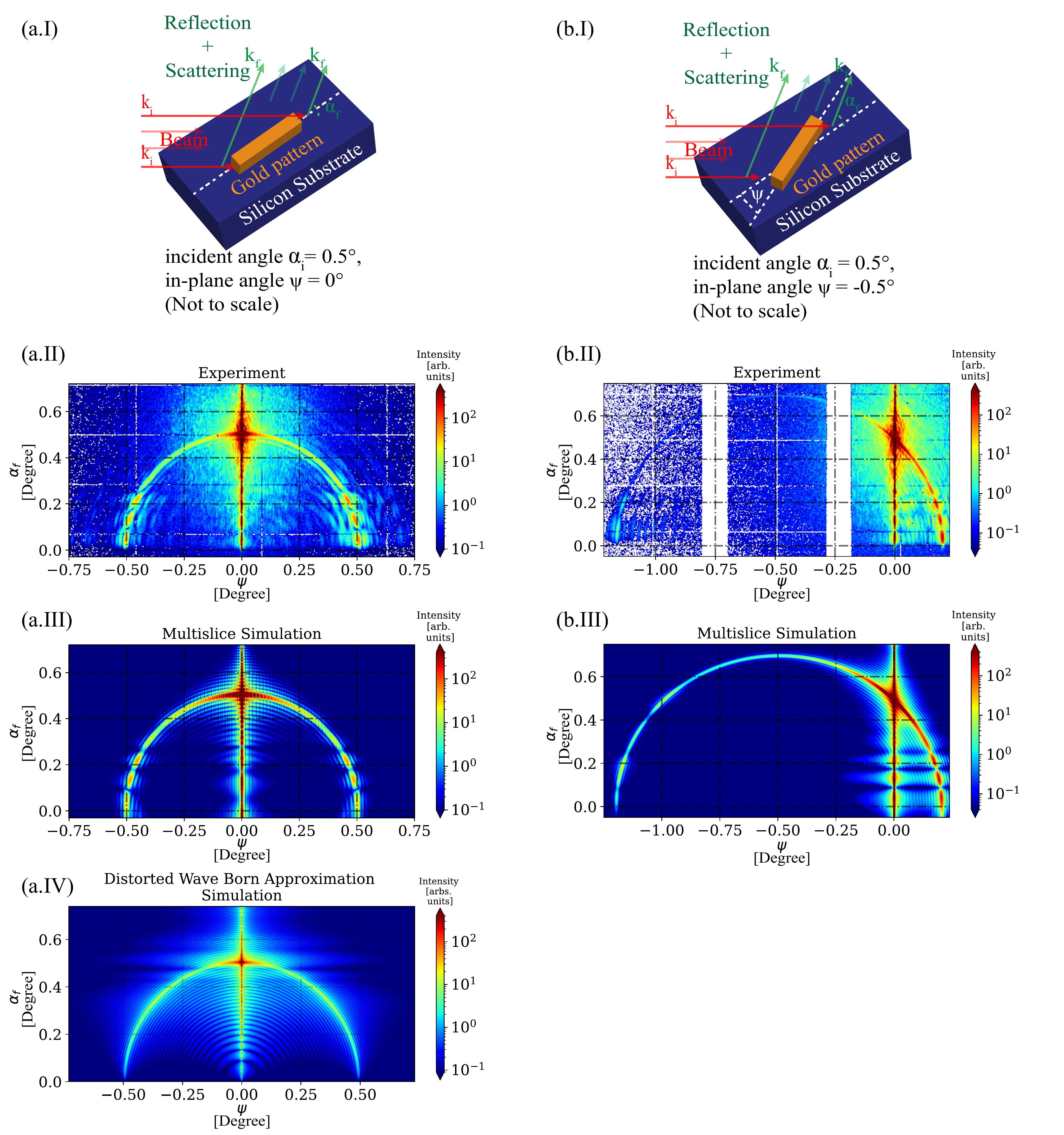}
\caption{Elongated Rod A pattern scattering images and simulations. \textbf{Left column (a)}: The rod is at an X-ray incident angle of 0.5$^\circ$ with zero in-plane rotation angle, schematically potrayed in \textbf{(a.I)}. \textbf{Right column (b)}: the rod is at the same incident angle of 0.5$^\circ$ with 0.5$^\circ$ in-plane rotation angle, as portrayed in \textbf{(b.I)}. The gaps in the plot \textbf{(b.II)} are due to modular gaps on the detector, whereas gaps are patched in the plot \textbf{(a.II)} by taking two images at offset detector locations. In both cases, multislice model is able to exactly reproduce the experimental scattering images as seen in \textbf{(a.III)} and \textbf{(b.III)}. The plot \textbf{(a.IV)} shows what DWBA simulations is unable to reproduce the fringes from dynamical scattering below $\alpha_f$ $\sim$ 0.2$^\circ$.}
\label{fig:PetraIIIAprime}
\end{figure}

Multislice simulations of the simple elongated rod A agree with experimental scattering images, which showing dynamical scattering, or beatings, near sample horizon as seen in Figure \ref{fig:PetraIIIAprime}. The X-ray incident angle was chosen to be 0.5$^\circ$ to highlight dynamical scattering phenomenon. It could be noted that the multislice simulation does not have a sharp drop of scattering intensities at the sample horizon ($\alpha_i = 0$) as seen in experiments. This is due to the fact that multislice propagation does not have enough empty Silicon substrate (due to GPU memory limits) to attenuate intensities below 
$\alpha_i = 0$. Additionally, the propagation past the gold pattern does not have a significant influence on the scattering intensities above the sample horizon. In Figure \ref{fig:PetraIIIAprime} a.IV, a DWBA simulation is also shown, in which dynamical scattering fringes cannot be reproduced. These fringes are dependent on the thickness of the samples and can be accurately predicted by multislicing simulations as also confirmed by SEM or AFM. Multislice is again shown to give accurate scattering patterns when the sample is rotated in the beam, for example, in-plane rotations up to $\psi$ = 0.5$^{\circ}$, as shown in the right column of Figure \ref{fig:PetraIIIAprime}. This will be useful for the reconstruction of tomography experiments in order recover the limited fieldview along the footprint direction due to the limited field of view in that direction. In all multislice simulations, scattering intensities can also be matched to experimental values by using the incident X-ray probe with a known photon flux. 

\section{Reconstruction using the multislice forward model}
\label{sec:Reconstruction using the multislice forward model}
In utilizing the multislice formalism for image reconstruction purposes, we can consider two regimes: kinematical regime, which is for scattering experiments done at angles well above the critical angles (e.g. 0.7$^{\circ}$ and above for samples made of silicon and gold is where dynamical scattering starts to get weaker) and dynamical regime, which is for those done at angles between 0.3$^{\circ}$ and 0.5$^{\circ}$. Up to now, most CDI experiments are performed in the forward transmission geometry and thus reconstruction algorithms adopt the kinematical approximations. The multislice model can be used to deal with both kinematical and dynamical regimes, but the reconstructions here are done in the complex dynamical regimes because the dynamical scattering phenomena can be only simulated by using multislice as discussed in Section \ref{sec:Fast Computing Multislice Simulations in GPU  and Experimental validation}. In the first subsection, the formalism is used to reconstruct a buried 3D pattern arrangement by minimizing a cost function between a ground truth scattering pattern and PyTorch's guesstimated scattering pattern: a simulated toy model reconstruction of buried chip pattern. In the second subsection, a single-shot experimental scattering image of the 3D FIB sample is used to perform 3D structural refinement using PyTorch's optimization, where measurements from Atomic Force Microscopy are used as a strong object support to help with PyTorch's optimization.

\subsection{CSSI-CDI reconstruction from simulation data: a simulated 3D pattern buried within a silicon substrate}
\label{subsec:CSSI-CDI reconstruction from simulation data: a simulated 3D pattern buried within a thick silicon substrate}
 
Reconstructions from scattering images acquired near the critical angle are challenging due to dynamical scattering. A model that can help explain dynamical scattering is a critical component of any reconstruction model. To do the reconstructions near the critical angle, a simple Mean Square Error (MSE) cost function could be written as described in Equation \ref{equ:loss-function} such that the difference between the ground truth (experimental scattering amplitude data or in this scenario simulated scattering ground truth amplitude, i.e. simulated scattering applied with poisson noise and rounded to mimic a physical detector) and an educated guess from PyTorch that is iteratively updated. Such a cost function could be minimized using PyTorch's automatic differentiation, updating parameters by using gradients computed by PyTorch. In other words, it is intrinsically a multiple parameter optimization problem, in which parameters define pattern's shape. By iteratively updating the guessed object used for the multislice simulation and reducing the cost function of difference between ground truth and guessed object, the buried 3D object can be reconstructed.

\begin{eqnarray}
\textrm{Loss}(T_{\textrm{predicted}} - T_{\textrm{ground truth}})=\frac{1}{N} \sum _{i=1}^{N = \textrm{Total pixels}}\left(T_{\textrm{predicted}}^i-T_{\textrm{ground truth}}^i\right)^{2}
\label{equ:loss-function}
\end{eqnarray}

\begin{figure}[!ht]
\centering\includegraphics[width=13.3cm]{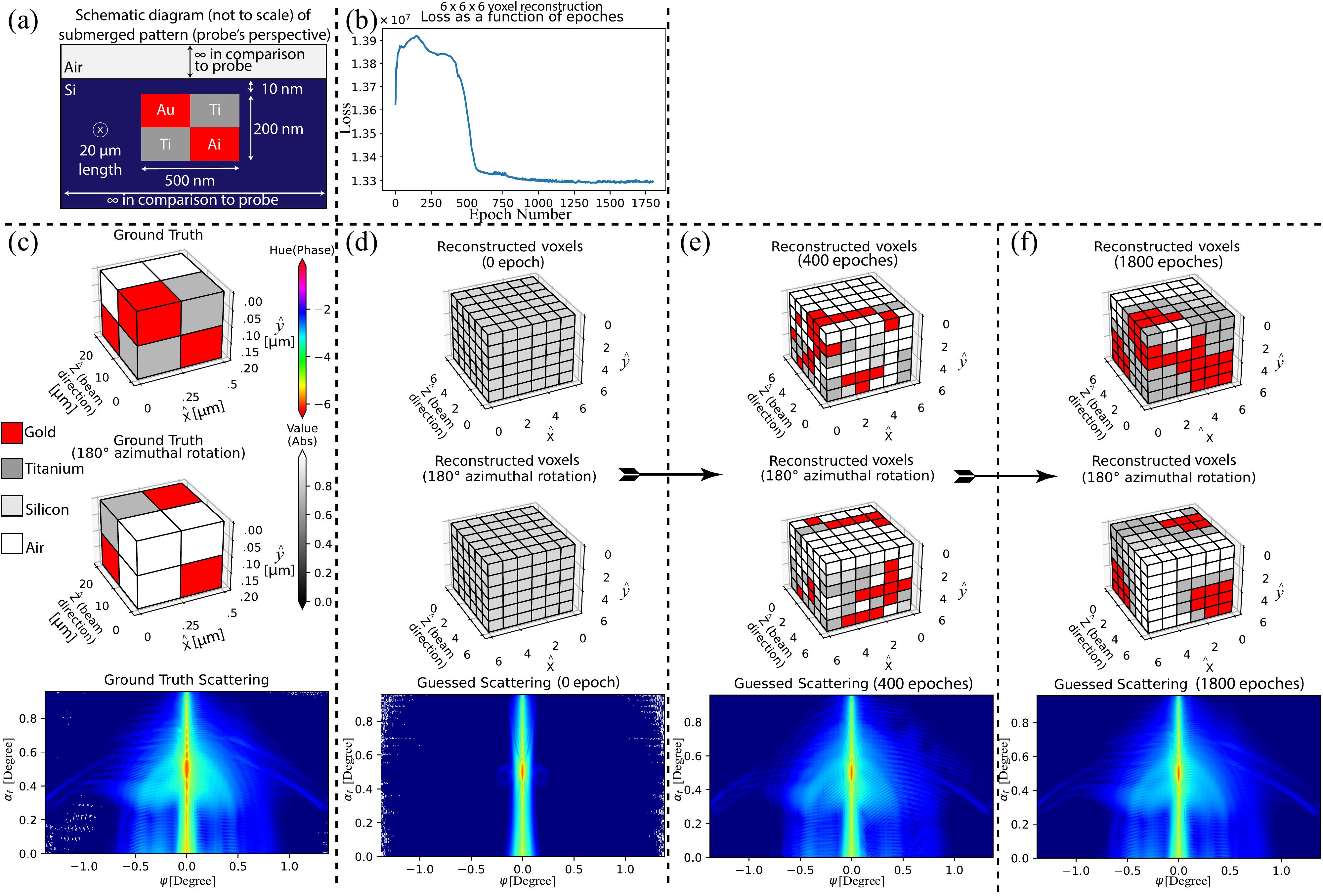}
\caption{Reconstruction of a buried 3D structure inside a Silicon substrate. \textbf{The leftmost column} is for the ground truth, detailing a 2D cross-section \textbf{(a)}, the 3D structure itself, and the computed ground truth scattering \textbf{(c)}, which is applied with poission noise and rounded to nearest integer value to emulate a detector image. \textbf{(d, e, f)} describe the 3D object being updated and its corresponding scattering images at epoches 0, 400, and 1800. \textbf{(b)} shows the loss function as a function of epoch, showing convergence to lowest cost function value to match the ground truth scattering.}
\label{fig:buried_chip}
\end{figure}

A more complicated cost function could be written such that 216 parameters are used to define 216 ( 6 x 6 x 6) voxels which describe a buried pattern arrangement (or rather a simple chip pattern with voxel's pixel resolution as the resolving limit of the imaging technique). Each parameter describes whether a voxel is just silicon (if all ten parameters are silicon, we have reflection from a bare silicon) or gold. A ground truth is first picked: a particular CSSI scattering pattern (with poisson noise) from a 3D buried chip. Then, the cost function is minimized and the parameters are updated; the difference between ground truth scattering pattern and the guessed scattering patterns is reduced. The reason why the cost function is written using scattering amplitude is because in actual image reconstructions, we only have scattering intensities with lost phase information. The optimization is also written such that 216 parameters that define voxels are applied with a filter after 20 epoches of optimization by taking nearest approximation to refractive indexes of interest; in other words, voxels are represented by integrated refractive index indicating a mixture of air, silicon, and/or gold, etc. For each iteration, an educated guess using gradients from automatic differentiation is made on 216 parameters using ADAM optimizer \cite{kingma2014adam} and a buried pattern arrangement is created and then multislice wave propagation is computed through the created object to obtain the final scattering pattern. 

The buried pattern arrangement is about 500 nm wide and 200 nm high, situated inside silicon, 10 nm below the surface. Each reconstruction voxel is about 10 $\mu$m (length) $\times$ 250 nm (width) $\times$ 100 nm (height). The voxels are elongated along the direction of the beam because of the small field of view along the wave propagation direction. It is important to note that these voxels are reconstruction voxels, not the multislice voxels, which are of much smaller dimensions as mentioned in Section \ref{sec:Fast Computing Multislice Simulations in GPU  and Experimental validation}. The sample is set to the X-ray incident angle of $\alpha_{i}$ = 0.5$^\circ$ . In the bottom leftmost of Figure \ref{fig:buried_chip}, the scattering image shown there is the ground truth that is used to minimize the cost function between itself and guessed scatterings for each epoch. Starting from a pure silicon substrate without features, the cost function minimization is observed to converge to ground truth after 1800 multislice propagation iterations or epoches. Different initial guessed conditions were also tested and they were all observed to converge to the ground truth. This simple 216 parameter optimization shows that multislice simulations are sensitive to the complex dynamical scattering near the critical angle and it is possible to reconstruct a three-dimensional buried pattern from a single scattering pattern if coherent X-ray flux and the field of view are sufficiently high and there is sufficient computing resources to support the high resolution reconstruction.

\subsection{CSSI-CDI structural refinement from experimental data: a 3D pattern created by Focused Ion Beam.}
\label{subsec:CSSI-CDI structural refinement from experimental data: a 3D pattern created by Focused Ion Beam}

\begin{figure}[!ht]
\includegraphics[width=13.3cm]{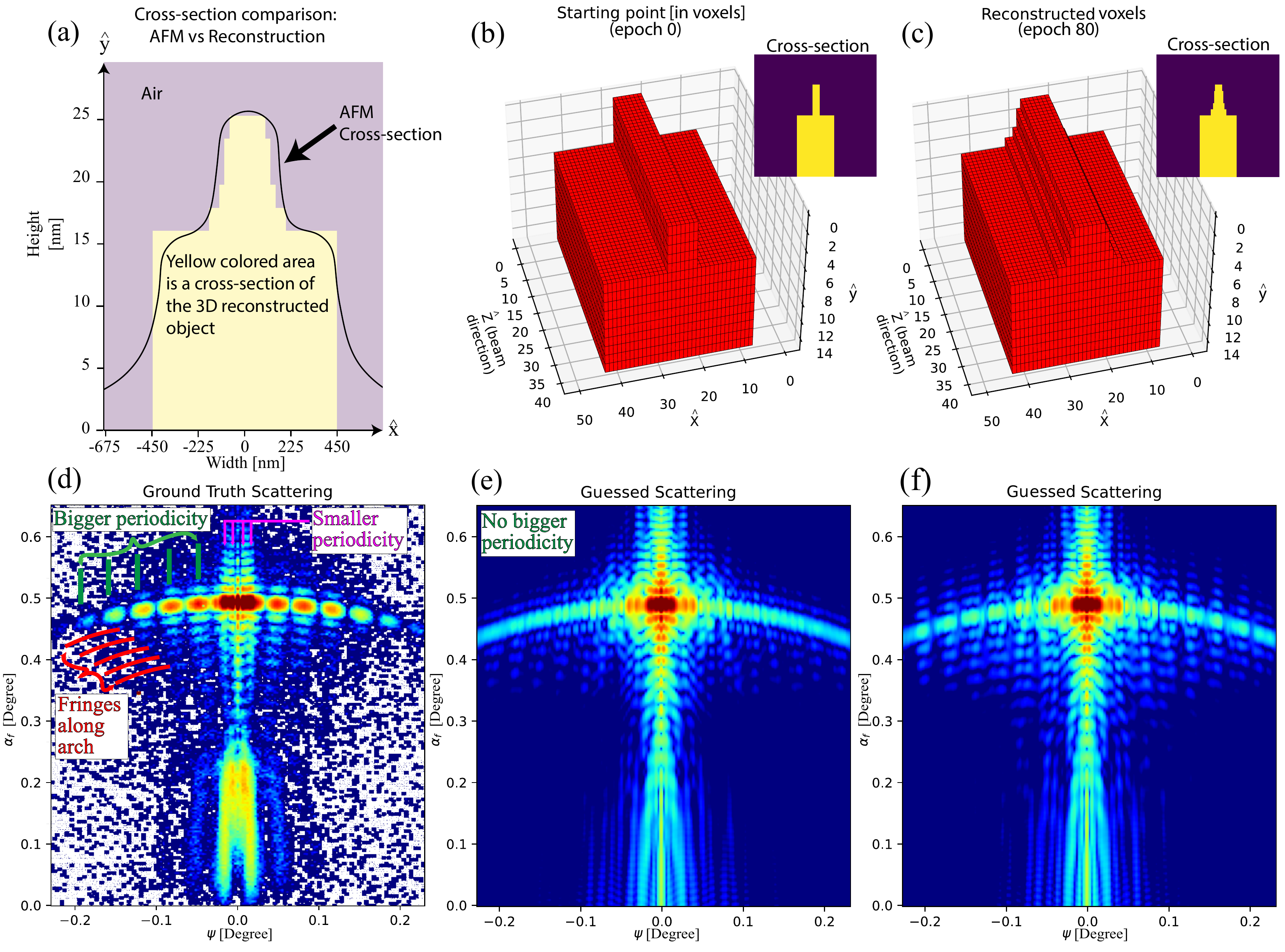}
\caption{a 3D CSSI-CDI structural refinement from a single shot experimental scattering pattern using prior knowledge. \textbf{(d)} shows ground truth scattering image obtained from the experiment and \textbf{(a)} the measured AFM cross-section profile juxtaposed on top of refined PyTorch's 3D structure. \textbf{(b, c)} show the initial starting point which is an idealistic 3D structure and its dimensions were from AFM measurements. The guesstimated scattering intensities in \textbf{(e)} have similarities with the ground truth, but the bigger periodicity (green annotations on ground truth scattering) in the scattering is missing. \textbf{(c, f)} show structurally refined 3D structure along X-Y plane after 80 epoches, where tapering profile at the top is observed to be responsible for the bigger periodicity in the guesstimated scattering pattern \textbf{(f)}. The smaller periodicity is present for both epoch 0 and 80 and matches with the ground truth scattering since it is due to bigger lengthscale, i.e. bottom 0.9 $\mu$m wide structure.}
\label{fig:Fib_expfit}
\end{figure}

In this subsection, parameter optimization utilized in the previous subsection \ref{subsec:CSSI-CDI reconstruction from simulation data: a simulated 3D pattern buried within a thick silicon substrate} is applied to an experimental scattering image. The experimental scattering image of the FIB 3D structure from Table \ref{tab:samplelist} was taken with X-ray flux of $5 \times 10^{9}$ Photons s$^{-1}$ where as for the buried 3D structure in \ref{subsec:CSSI-CDI reconstruction from simulation data: a simulated 3D pattern buried within a thick silicon substrate}, the simulated X-ray flux was 100 times more. The Advanced Photon Source is going through an upgrade that will increase its coherent by two orders of magnitude, but currently the experimental scattering image of the FIB 3D structure does not have enough X-ray flux to resolve finer lengthscales along $\hat{y}$(height) and $\hat{z}$(length) directions (i.e. numerical aperture is flux and geometry limited in a single exposure setup). Therefore a strong object support or prior knowledge could be used as a guide to do fine-tuning or structural refinement using the PyTorch optimization. The CSSI geometry has asymmetric resolutions along each dimensions and the AFM image of the FIB sample has high sensitivity for height ($\hat{y}$) and length ($\hat{z}$), but has limitations along width ($\hat{y}$) especially at sharp edges due to the dull AFM tip profile. Given the constraints, it is demonstrated here horizontal beatings seen in the experimental X-ray scatterings are due to the tapering profile of the top layer of the FIB pattern — one-dimensional parameter optimization or structural refinement.  

In Figure \ref{fig:Fib_expfit}, the ground truth scattering from experiment is shown on the left bottom plot. It is important to note that there are smaller beatings noticed at small $\psi$ angles ($\sim 0.1^{\circ}$) or indicated by pink annotation on \textbf{d} of Figure\ref{fig:Fib_expfit}, which represent bigger lengthscales (i.e. the bottom part of the FIB object which is 0.9$\mu$m). The bigger beatings are indicated by green annotations and it will be discussed below that they are not entirely due to the smaller top part of the FIB object, which is 0.3$\mu$m. Fringes along arch (annotated by red texts and drawings) are due to both lengthscales along $\hat{z}$ and $\hat{y}$.   To simulate the entire FIB sample with reasonable resolution, 50 x 40 x 14 voxels (28000 parameters) are used to define the 3D shape of the FIB sample and up to 77 GB of GPU memory is occupied during forward calculations and gradient updates;  each voxel is about 600 nm (length) $\times$ 7 nm (width) $\times$ 1 nm (height) and each epoch, including forward calculation and gradient update, takes a total of 2.5 s for this sample size. If we start from a perfect two-layers as described in Table \ref{tab:samplelist}, the computed scattering image (bottom middle plot of the figure) does not have any bigger beatings as seen in experiment. We do not have enough lengthscales sensitivity along $\hat{y}$ because the resolution necessary to resolve the details of heights is restricted by vertical numerical aperture which in turn is limited by the X-ray flux for a single exposure reconstruction. Additionally the information along beam $\hat{z}$ is also limited due to X-ray flux issues. Therefore,  the initial starting point for structural fine-tuning is best set with perfect rectangles with heights, length, and width values given by the AFM image. One reason for starting with sharp edges, rather than using AFM profile is to retrieve spatial variation along $\hat{x}$ direction without being influenced by the AFM tip convolution issues. The optimization is carried out, using the MSE loss function and the ADAM optimizer. Every twenty epoches, a shrink wrap support \cite{marchesini2003x} is applied, which also acts as nearest neighbor filter to determine whether a voxel is platinum or air. After 80 epoches, the guessed scattering starts to show the bigger beatings (along the arch) which agree with the experimental ground truth. This confirms that the tapering profile of the top layer of the FIB sample contribute to the bigger beatings. The goal here is to demonstrate that thousands of parameters can be updated per each epoch through fast computations and optimization in GPU, as long as the experimental data allows it. In this case, there is only enough flux for physical information along the $\hat{x}$ direction, requiring strong sample support information from AFM measurements for height($\hat{y}$) and length($\hat{z}$). Higher X-ray fluxes, which will be available in Advanced Photon Source Upgrade, will enable better sensitivity to the 3D structure, as demonstrated by the simulation exercise in Subsection above. Since distributed memory algorithms for X-ray wave propagation have been implemented \cite{ali2020comparison}, the next critical step is to implement this algorithm across multiple GPUs to enable experimental reconstructions of larger lengthscales at higher resolution, which require large fieldview or are memory intensive. 

\section{Conclusion}

Coherent Surface Scattering Imaging (CSSI) utilizes the advantages of the coherent diffractive imaging concept to reconstruct coherent grazing-incidence scatterings from surface and thin-films structures. High-flux scattering images taken for tomography will further complement CSSI for full 3D reconstructions with isotropic high-resolution in all directions. Whatever solutions there may be in the instrumentation aspect of experiments, multislice is crucial in providing a holistic forward model in the CSSI imaging technique. One big advantage of the reflection-geometry multislice approach is that it can be applied to any three-dimensional object structure with inhomogeneous refractive index distribution $n(x, y, z)$ and any incoming X-ray probe shape and phase as long as a sufficient computational resource (memory, computing nodes, etc) makes it possible. Additionally, multislice model can simulate ptychography and tomography simulations in both dynamical and kinematic regimes without requiring the plane wave assumption as in DWBA and without any restrictions on the form of the X-ray probe nor its phase. Unlike DWBA theory only applicable for far field scattering analysis, multislice as a wave proportion method is also capable for near-field imagining analysis. The next critical step is therefore to implement automatic differentiation and computation of the multislice model across multiple GPUs to simulate with larger field view and enable experimental reconstructions of larger sample sizes and smaller voxel sizes.

With continuing advances in synchrotron X-ray sources, the Coherent X-ray Scattering Imaging technique will be able to probe smaller lenghthscales at shorter timescales. Resolution to differentiate layer by layer, or in other words depth sensitivity, can be achieved in experiments by varying X-ray incident angles and changing in-plane rotations. Such experimental implementations along with the multislice forward model open the door to myriads of imaging techniques such as CSSI-CDI (reconstruction from a single shot scattering image from a small sample), CSSI-Ptychography (reconstruction from scattering images of overlapping scans on an extended sample), and CSSI-tomography (reconstructions from scattering images of sample at different in-plane angles), and CSSI-laminogrphy (combination of tomography and ptychography for 3D reconstruction of an extended object).

\section{Backmatter}

\begin{backmatter}
\bmsection{Funding}
This work is supported by the Advanced Photon Source, a US Department of Energy (DOE) Office of Science User Facility. Fabrication work was performed at the Center for Nanoscale Materials, a U.S. Department of Energy Office of Science User Facility. Both Advanced Photon Source and Center for nanoscale Materials are supported by the U.S. DOE, Office of Basic Energy Sciences, under Contract No. DE-AC02-06CH11357. P.M., A.T., and Z.J. are supported by the DOE Early Career Research Program. 

\bmsection{Acknowledgments}
P.M., M.C. and Z.J. took the X-ray experimental data. P.M. developed CuPy and PyTorch codes to do forward model calculations and reconstructions based on M.C.'s initial work on multislice simulations. M.J.W made the rod A sample with e-beam lithography.  J.Z. made the Focused Ion Beam sample and performed the Atomic Force Microscopy and Scanning Electron Microscopy measurements on it. All the other co-authors were involved in experiments,  discussions, and in the preparation of the manuscript. 

\bmsection{Disclosures}
The authors declare no conflicts of interest.

\bmsection{Data availability} Data underlying the results presented in this paper are not publicly available at this time but may be obtained from the authors upon reasonable request.

\end{backmatter}
\bibliography{multislice}

\end{document}